\documentclass[pre,twocolumn,showpacs]{revtex4}

\usepackage{amsmath}
\usepackage{amsfonts}
\usepackage{bm}
\usepackage{epsfig}

\begin{document}

\title{Coupling of thermal and mass diffusion in regular binary
thermal lattice-gases} 

\author{Ronald Blaak}
\email{rblaak@pa.uc3m.es}
\affiliation{Grupo Interdisciplinar de Sistemas Complicados (GISC),
Departamento de Matem\'aticas, Universidad Carlos III de Madrid,
Avda. de la Universidad, 30, 28911, Legan\'es, Madrid, Spain}

\author{David Dubbeldam}
\email{dubbeldam@its.chem.uva.nl}
\affiliation{Department of Chemical Engineering, University of
Amsterdam, Nieuwe Achtergracht 166, 1018 WV Amsterdam, The
Netherlands}

\date{\today}

\begin{abstract}
We have constructed a regular binary thermal lattice-gas in which the 
thermal diffusion and mass diffusion are coupled and form two
nonpropagating diffusive modes. The power spectrum is shown to be
similar in structure as for the one in real fluids, in which the
central peak becomes a combination of coupled  entropy and
concentration contributions. Our theoretical findings for the
power spectra are confirmed by computer simulations performed on this 
model. 
\end{abstract}

\pacs{05.20.Dd, 06.50.+q, 05.60.-k}

\maketitle

The power spectrum of light scattered by a binary solution is 
more complicated than that of a single component fluid. The central peak
contains combined effects of entropy and concentration fluctuations
\cite{Mountain,BoonYip}. The cross effects are well known in
nonequilibrium thermodynamics as the Dufour effect and the Soret
effect, and are caused by the coupling between heat-flow and
diffusion. For example, heat may be transported by conduction but also
by diffusion of the two components. In the simplified two-dimensional
model presented here this coupling phenomena can be analyzed in detail.

Lattice gas automata (LGA) are plagued by many defects. Although some
of these defects can be solved, the models are in general not suitable
for modeling realistic fluids because they do not exhibit a fully
realistic thermodynamical behavior. They are, however, useful tools
for understanding more fundamental problems in thermodynamics related
to discretization and testing concepts in  
statistical mechanics \cite{Boon2001}. In LGA the positions and
velocities of point-like particles are discretized onto a
lattice~\cite{Fri86}. The dynamics consist of a cyclic process
of propagating particles according to their velocity to a
neighboring node followed by local collisions that typically conserve
mass and momentum. 
 
In order to recover the macroscopically isotropic Navier-Stokes
equations  the lattice in two-dimensions is usually chosen to be
hexagonal. Apart from problems related to this, early LGA were also
plagued by unwanted spurious
invariants~\cite{Brito:1991JSP,Das:1992PA,Ernst:1992JSP}. 
A satisfactory extension to include thermal properties in LGA 
 was made by Grosfils, Boon, and Lallemand
(GBL), who introduced a multiple speed model, defined on a
two-dimensional hexagonal lattice~\cite{GBL92}.  The model uses a
velocity set consisting of a single rest particle and three 
rings, each containing six directions, with velocities of magnitude $1$,
$\sqrt{3}$, and $2$. 

Generalizations of LGA to mixtures have been used to study interfaces
and phase transitions~\cite{Rothman:1994RMP}. Some of these models use
a passive label to distinguish the 
different species~\cite{Chen:1987PF}. In other models the particles live
on different lattices~\cite{Dab:1991PRL,Kapral:1991PRL}. However,
these models are athermal and do not show the coupling phenomena we are
interested in. A two species thermal model by using a passive
label, is also not suitable, because mass and heat transport would
completely decouple~\cite{Blaak:2001PRE}.

Here we introduce, for the first time, a thermal binary lattice-gas
model capable of capturing the essence of a real mixture with respect
to coupling of entropy and concentration fluctuations. 
This model allows us to calculate the coupling
quantitatively and separate the different contributions, which in the
continuous case in general only is possible in the low density
limit. The two species of particles live on separate two-dimensional
hexagonal lattices and are labeled red and blue. Independently they
would behave as normal GBL models, but we allow the particles to
interact during the collision phase, i.e. momentum and/or energy can
be transferred from one to the other lattice. The number of red and
blue particles, however, is constant in every collision. 

The state of a node can be specified by a set of Boolean occupation
numbers $n_{i\mu}$, denoting the  presence or absence of a particle of
type $\mu=\{r,b\}$ in velocity channel $c_i$, where $i$ is a label
running over all 19 velocities. Due to the Boolean nature of LGA,
the ensemble average of the occupation numbers $f_{i\mu}$ in
equilibrium, is described by a Fermi-Dirac distribution
\begin{equation}
\label{Eq:FermiDirac}
f_{i\mu} \equiv \langle n_{i\mu}\rangle = \frac{1}{1 + e^{- \alpha_\mu 
+ \beta \bm{c}_i^2/2  -\bm{\gamma} \cdot \bm{c}_i}},  
\end{equation}
where $\alpha_r$, $\alpha_b$, $\beta$, and $\bm{\gamma}$ are
Lagrange multipliers. Here, $\beta$ is the inverse temperature,
$\alpha_r$ and $\alpha_b$ fulfill a chemical potential role, and
${\bm \gamma}$ is a parameter conjugate to the flow velocity. For
simplicity we will work in the overall zero momentum case by setting
$\bm{\gamma}=0$. 

A collision outcome is chosen with equal probability amongst all
members of an \emph{equivalence class}, i.e. a group of states having
the same red mass, blue mass, total momentum, and total energy.
In a \emph{regular binary mixture} at most two particles, each  of
different type, can be in the same velocity
state~\cite{Ernst:1990JSP}. This differs from the \emph{color
mixture} where the particles of the original single specie model are
given a color to distinguish between them and was analyzed in the case
of the GBL color mixture (CGBL) \cite{Blaak:2001PRE}. 
In that case the collision operator could be split into two separated
steps: a GBL collision and an independent redistribution of the
colors, which made the simulations a relative simple extension of
normal GBL simulations, even though the number of different states in
the CGBL model was $3^{19}$. Here we have a ``true'' $38$-bits 
model leading to $2^{38}$ states on which the collision operator has
to act. Clearly a naive lookup-table strategy in order to simulate
this system is only feasible if the table is constructed partially and
stored temporarily during the simulations, due to the excessive memory
requirements for storing the complete table.

Regular binary mixtures, however, do allow for a convenient solution,
which is less efficient than storing the complete collision table, but
has still a relative good performance (about a factor 2-5 slower than
CGBL). Rather than storing a collision table based on the states, we
make one based on the different classes. This allows us to generate a
set of outgoing classes for the two species and via a GBL-like
process an outgoing state. If this is combined with the 48
symmetry operations (six rotations, two reflections, red/blue
symmetry, and particle/hole symmetry), we get a working algorithm that
can be used on a computer with 256 MB of memory~\cite{Blaak:2001JSP}. 

The theoretical framework for thermal lattice-gases is well
established \cite{GBL93}, and is here extended to binary mixtures. 
It is possible to solve the
many-body dynamics of LGA  by using the Boltzmann molecular chaos
assumption. Although one would expect deviations for higher densities,
it turns out that these deviations are very small, in fact for many
purposes even within a few percent accuracy. Therefore, provided the
fluctuations in the average occupation numbers are small, a Taylor
expansion of the collision term in the neighborhood of the equilibrium
distribution is justified, yielding a linearized collision operator
$\Omega$. Then, in first approximation the behavior of the system can
be obtained by analyzing these deviations in terms of eigenmodes. 

In the analysis of the behavior of fluctuations in LGA mixtures it is
convenient to introduce the colored scalar 
inproduct~\cite{Blaak:2001PRE,Hanon}
\begin{equation}
 \langle A|B\rangle=\sum_{i\mu}A({\bm c}_{i\mu})B({\mathbf
 c}_{i\mu}) \kappa_{i\mu},
\end{equation}
where $\kappa_{i\mu}=f_{i\mu}(1-f_{i\mu})$ is the variance in the
average occupation number. For reasons of symmetry $\kappa$ is
included in definition of the right vectors, i.e. $|B\rangle_{i\mu} =
\kappa_{i\mu} B(\bm{c}_{i\mu})$. 

Following the method of R\'esibois and Leener \cite{Resibois} we need
to find the ${\bm k}$-dependent eigenfunctions and eigenvalues of
the single-time step Boltzmann propagator
\begin{equation}
\label{Eq:eigenproblem-right}
e^{-\imath \bm{k}\cdot \bm{c}}(\bm{1}+\Omega)
|\psi(\bm{k})\rangle = e^{z(\bm{k})}| \psi(\bm{k})\rangle.
\end{equation}
Within this formulation the hydrodynamic modes are characterized by 
the fact that the eigenvalues $z_\mu$ should go to zero in 
the limit of small wavevectors. Therefore we can derive these ${\bm 
k}$-dependent eigenvectors and eigenvalues by making a Taylor
expansion in $\bm{k}$. The resulting eigenvalues to second order in
${\bm k}$ are given by 
\begin{eqnarray}
 z_\pm({\bm k})&=&\pm\imath c_s{\bm k}-\Gamma {\bm k}^2,\\
 z_\perp({\bm k})&=&-\nu {\bm k}^2,\\
 z_{s_\pm}({\bm k})&=&-s^\circ_{\pm} {\bm k}^2.
\end{eqnarray}
It is with these modes that the binary-mixture responds to deviations
from thermal equilibrium. The first two eigenmodes describe sound
propagating (in opposite directions parallel to ${\bm k}$) with
$\Gamma$ the sound damping coefficient and $c_s$ the adiabatic sound
speed.  The third eigenvalue describes a shear mode with $\nu$ the
shear viscosity coefficient. The last two eigenvalues represent
purely diffusive, non-propagating processes, but are in general not
directly related to a familiar transport coefficient. Rather they
always appear in combination with each other and can be expressed in
other transport properties by 
\begin{equation}
s_\pm^\circ= \mbox{$\frac{1}{2}$}(\chi+ {\cal D}) \pm
\mbox{$\frac{1}{2}$} \sqrt{4 Q^2 + (\chi-{\cal D})^2}, 
\label{Eq:smodes}
\end{equation}
where $\chi$ is the generalized thermal diffusivity, ${\cal D}$ a
property related to the mass diffusion, and $Q$ can be considered as a
measure of the correlation between the two. Another interesting
relation can be obtained by introducing the ratio of the specific
heats $\gamma$
\begin{equation}
\Gamma=\mbox{$\frac{1}{2}$} \left(\nu+(\gamma-1)\chi\right).
\end{equation}

\begin{center}
\begin{figure}[t]
\epsfig{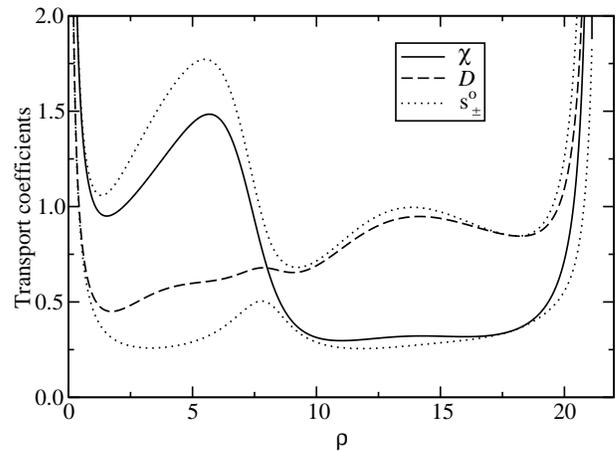}
\caption[a]{
The diffusive transport values as a function of the density
at reduced temperature $\theta=0.1$ and fraction of red particles
$P_r=0.9$.}
\label{Fig:coupling} 
\end{figure}
\end{center}

In Fig.~\ref{Fig:coupling} the diffusive transport properties are
shown at fixed reduced temperature $\theta=\exp(-\frac{1}{2}\beta)$.
The modes $s^\circ_\pm$  converge for low densities to the thermal
diffusivity and mass diffusion, due to the decoupling of the
fluctuations in entropy and energy. This behavior resembles the
situation for binary solutions where a formula similar to
(\ref{Eq:smodes}) exists \cite{BoonYip}, with the same decoupling in
the very dilute limit. In our model this means that the ratio ${\cal
Q}/(\chi - {\cal D})$ vanishes. The value of ${\cal Q}$, however, will
in general remain small but finite due to the divergencies of the
transport coefficients in the low density limit of LGA.

The decoupling is also observed in the limit of a single
specie, i.e. the fraction of blue particles $P_b$ goes to zero. In
this limit the model reduces to a normal GBL model, albeit that the
diffusion related property ${\cal D}$ will remain finite. For reasons
of symmetry $s_\pm^\circ$ also converge to $\chi$ and ${\cal D}$ in
the limit of equal red and blue density. In fact this special limit
can be analyzed completely in a similar way as done for
CGBL~\cite{Blaak:2001PRE}. Finally decoupling appears in two
other cases due to the duality of the model under interchanging
particles and holes, these are the high density limit and the limit in
which one of the sublattices is almost completely filled.

The decoupling arises from an effective equipartition in the model
making the ratio of the average occupation numbers $f_{ir}/f_{ib}$
the same constant for each velocity channel $i$. As can be seen in
Fig.~\ref{Fig:coupling} there seem to be also intermediate values at
which decoupling appears. Although indeed the $s_\pm^\circ$ converge
to $\chi$ and ${\cal D}$ this is not due to decoupling via
equipartition but due to the cancellation of terms. This property could
obviously be used in the analysis, but the location of
these points depends in a non-trivial way on the chosen 
system parameters. 

The Boltzmann approximation enables us to calculate the modes up to
fairly large wavevectors, even within the generalized
hydrodynamic regime. In the hydrodynamical regime of small wavevectors
${\bm k}$, small frequencies $\omega$, the hydrodynamical modes are
well separated from the kinetic modes that, due to their exponential
decay, can be neglected. In combination with a Taylor-expansion, this
allows one to derive a Landau-Placzek approximation of the spectral
density $S({\bm k},\omega)$ 
\begin{equation}
 \begin{split}
 &\frac{S({\bm k},\omega)}{S({\bm k})} =\sum_\pm
  \frac{\gamma-1}{\gamma}
     \left(1\pm\frac{\chi-{\cal D}}{s^\circ_+-s^\circ_-}\right)
     \frac{s^\circ_\pm k^2}{\omega^2+(s^\circ_\pm k^2)^2}\\
 &\qquad +\sum_\pm\frac{1}{\gamma}\frac{\Gamma k^2}{(\omega\pm c_s
 k)^2+(\Gamma k^2)^2}\\ 
 &\qquad+\frac{1}{\gamma}\left[\Gamma+(\gamma-1)\chi\right]\frac{k}{c_s}
   \sum_\pm\frac{c_s k\pm\omega}{(\omega\pm c_s k)^2+(\Gamma k^2)^2},
  \end{split}
\label{Eq:FullLP}
\end{equation}
where $S({\bm k})$ is the static structure factor.

The spectrum contains an unshifted central peak and is formed by two
Lorentzians due to the non-propagating processes characterized by
$s_\pm^\circ$. The shear mode does not contribute to the spectrum and
the two propagating modes lead to the presence of the two frequency-shifted
Brillouin lines.  The last two terms in Eq.~(\ref{Eq:FullLP}) have a
contribution orders of magnitude smaller than the amplitude of the
Lorentzians and induce a slight pulling of the peaks towards the
central peak~\cite{BoonYip}. The symmetry of the different
contributions is such that the ratio of the integrated contributions
of the central peak and the Brillouin components is constant and given
by $\gamma-1$. 

\begin{center}
\begin{figure}[t]
\epsfig{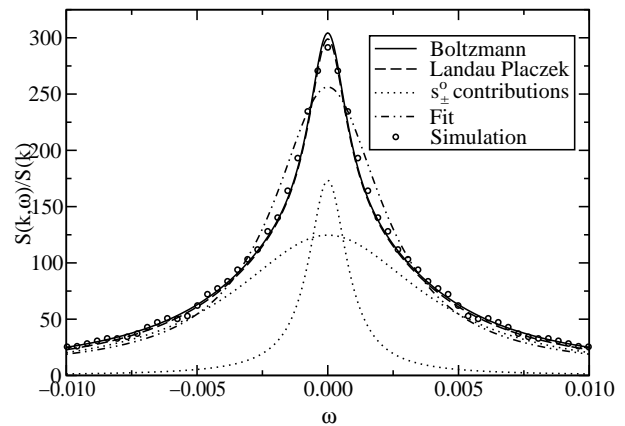}
\caption[a]{
The central part of the spectrum in the Boltzmann approximation, the
Landau-Placzek approximation, the $s_\pm^\circ$ contributions are
shown separately. The system parameters are $\theta=0.1$, $\rho=6.5$,
$P_r=0.05$, $k_x=4\times2\pi/512$.  The wavevector ${\bm k}$
and frequency $\omega$ are given in reciprocal lattice and time
respectively.}
\label{Fig:central_spectrum}
\end{figure}
\end{center}

Fig. \ref{Fig:central_spectrum} illustrates the composition of the
central peak. The simulation results overlap
perfectly with the Boltzmann and Landau-Placzek approximations.
The contributions of the diffusive modes to the central peak are
indicated separately and for comparison a least square fit based on
Eq.~(\ref{Eq:FullLP}) to the whole spectrum, including the Brillouin
peaks outside the interval shown, is made based on approximating the
central peak by a single Lorentz.

As can be seen from the Landau-Placzek expression (\ref{Eq:FullLP})
one of these contributions will vanish in the limits where
$s_\pm^\circ$ converge to $\chi$ and ${\cal D}$ and the central peak
reduces to
\begin{equation}
\frac{S^{\text{cen}}({\bm k},\omega)}{S({\bm k})} = 
\frac{\gamma-1}{\gamma}\frac{2\chi k^2}{\omega^2+(\chi k^2)^2},
 \label{Eq:limit}
\end{equation}
even though the other transport coefficient remains finite. As
mentioned before, this occurs at a limited set of locations, such as
the low/high density limit, single specie limits, and the limit of
equal red and blue density. In addition to these general limits it is
also found in the low/high temperature regions close to densities
where all rings of particles with the same absolute velocity 
 are completely filled or empty \cite{Blaak:2001PRE}.
This reduction of the central peak to a single Lorentz in the low
density limit is also found in the very dilute limit of binary
solutions. But contrary to what is found here, it is usually
the thermal diffusivity that disappears \cite{Mountain,BoonYip}.

If $s_\pm^\circ$ are sufficiently different, one can separate
the two contributions in the spectrum outside the limits mentioned
above (Fig.~\ref{Fig:central_spectrum}). This does not automatically
imply that one could obtain the transport values of interest from the
experimental spectrum only, because they can still differ
significantly from $\chi$ and ${\cal D}$. 

In the low density limit we can identify ${\cal D}$ with the
mass-diffusion of the GBL model. In the limit of a single specie this
is no longer true if one goes to higher densities, even though all
other properties converge to their GBL values. The origin of this
problem is found in the possibility to have more than one particle
with the same velocity in this model. Continuous theory suggests that
${\cal D}$ is not the correct generalization of the
diffusion~\cite{BoonYip}. This is consistent with the way in 
which this quantity arises in the present model~\cite{Blaak:2001JSP}. 

\begin{center}
\begin{figure}[t]
\epsfig{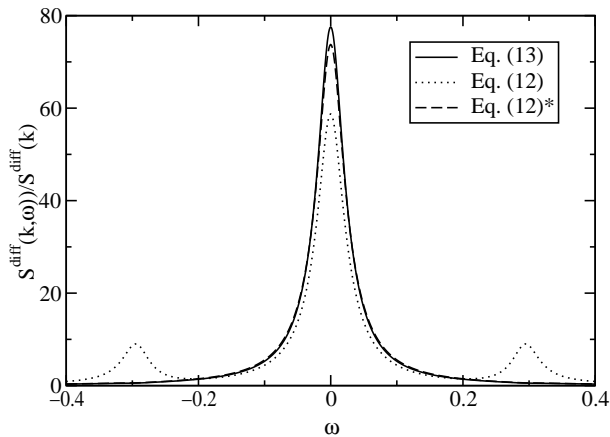}
\caption[a]{
A diffusive spectrum based on the different possible choices for the
density difference. The curve labeled by * is corrected by
subtracting the propagating part. The system parameters are $\theta=0.05$,
$\rho=10.0$, $\rho_r/\rho=0.15$, $k_x=20\times2\pi/512$.  The
wavevector ${\bm k}$ and frequency $\omega$ are given in reciprocal
lattice and time respectively.}
\label{Fig:diff_spectrum}
\end{figure}
\end{center}

The proper generalization of the diffusion is not completely
trivial. There is some ambiguity in the choice one needs to make. In a
continuous fluid one would consider the decay of a signal of the type
\begin{equation}
 |\text{diff}\rangle = \frac{|R\rangle}{P_r}-\frac{|B\rangle}{P_b},
\end{equation}
where $|R\rangle_{i\mu}=\delta_{r\mu}$ and
$|B\rangle_{i\mu}=\delta_{b\mu}$.
A spectrum based on this normalized density difference, however, leads to
Brillouin peaks as is shown in Fig.~\ref{Fig:diff_spectrum}. This was
already indicated earlier in an athermal binary mixture
\cite{Ernst:1990JSP}, where it was proposed to use 
\begin{equation}
 |\text{diff}\rangle=\frac{|R\rangle}{\langle R|R\rangle}
  -\frac{|B\rangle}{\langle B|B\rangle}. 
\end{equation}
But in the thermal case this signal suffers from the same problem. In
both cases one could in principle extract the propagating part, but it
is more natural to reformulate the density of interest to
\begin{equation}
 |\text{diff}\rangle=\frac{|R\rangle}{\langle p|R\rangle}
  -\frac{|B\rangle}{\langle p|B\rangle}, 
\end{equation}
with $p=c^2/2$ being the energy per particle. 
This problem originates from the lack of equipartition, but in general
the differences between the various choices will remain small
\cite{Blaak:2001JSP}.  

In conclusion, we presented a thermal binary mixture, which exhibits
spontaneous thermal and concentration fluctuations. The regular
mixture model is much more involved than the color mixture, but it is
closer to the dynamics of real binary mixtures, as it is able to
capture the coupling phenomena observed in binary solutions. The
coupling between 
energy transport and diffusion results in two diffusive non-propagating
modes and  in a more complicated structure of the power spectrum.
In general, however, these two ``true'' transport coefficients do not
seem to correspond with a conventional transport coefficient, but
always appear in combination with each other. 

The LGA spectrum is similar in structure as the one for the continuous
case. In general it is impossible to separate the entropy and
concentration contributions, and the central peak can not be described
by a single Lorentzian.
However in several limits, namely the
low/high-density limit, the single specie limit, and the equal
red/blue limit, the true modes $s_\pm^\circ$ converge to generalized
thermal diffusivity $\chi$ and a mass-diffusion like ${\cal D}$. 
In other cases the different transport coefficients can be determined
by using the theoretical framework, that provides an accurate
description as is demonstrated by the comparison of the Landau-Placzek
expression with simulation results. 

The authors would like to thank H.~Bussemaker and D.~Frenkel for
helpful discussions. R.B. acknowledges the financial support of the EU
through the Marie Curie Individual Fellowship Program (contract
no.~HPMF-CT-1999-00100).


\begin{thebibliography}{18}
\expandafter\ifx\csname natexlab\endcsname\relax\def\natexlab#1{#1}\fi
\expandafter\ifx\csname bibnamefont\endcsname\relax
  \def\bibnamefont#1{#1}\fi
\expandafter\ifx\csname bibfnamefont\endcsname\relax
  \def\bibfnamefont#1{#1}\fi
\expandafter\ifx\csname citenamefont\endcsname\relax
  \def\citenamefont#1{#1}\fi
\expandafter\ifx\csname url\endcsname\relax
  \def\url#1{\texttt{#1}}\fi
\expandafter\ifx\csname urlprefix\endcsname\relax\def\urlprefix{URL }\fi
\providecommand{\bibinfo}[2]{#2}
\providecommand{\eprint}[2][]{\url{#2}}

\bibitem[{\citenamefont{Mountain and Deutch}(1969)}]{Mountain}
\bibinfo{author}{\bibfnamefont{R.~D.} \bibnamefont{Mountain}} \bibnamefont{and}
  \bibinfo{author}{\bibfnamefont{J.~M.} \bibnamefont{Deutch}},
  \bibinfo{journal}{J. Chem. Phys.} \textbf{\bibinfo{volume}{50}},
  \bibinfo{pages}{1103} (\bibinfo{year}{1969}).

\bibitem[{\citenamefont{Boon and Yip}(1980)}]{BoonYip}
\bibinfo{author}{\bibfnamefont{J.~P.} \bibnamefont{Boon}} \bibnamefont{and}
  \bibinfo{author}{\bibfnamefont{S.}~\bibnamefont{Yip}},
  \emph{\bibinfo{title}{Molecular Hydrodynamics}}
  (\bibinfo{publisher}{McGraw-Hill Inc.}, \bibinfo{address}{New-York},
  \bibinfo{year}{1980}).

\bibitem[{\citenamefont{Rivet and Boon}(2001)}]{Boon2001}
\bibinfo{author}{\bibfnamefont{J.~P.} \bibnamefont{Rivet}} \bibnamefont{and}
  \bibinfo{author}{\bibfnamefont{J.~P.} \bibnamefont{Boon}},
  \emph{\bibinfo{title}{Lattice Gas Hydrodynamics}}, Cambridge Nonlinear
  Science Nr. 11 (\bibinfo{publisher}{Cambridge University Press},
  \bibinfo{address}{Cambridge}, \bibinfo{year}{2001}).

\bibitem[{\citenamefont{Frisch et~al.}(1986)\citenamefont{Frisch, Hasslacher,
  and Pomeau}}]{Fri86}
\bibinfo{author}{\bibfnamefont{U.}~\bibnamefont{Frisch}},
  \bibinfo{author}{\bibfnamefont{B.}~\bibnamefont{Hasslacher}},
  \bibnamefont{and} \bibinfo{author}{\bibfnamefont{Y.}~\bibnamefont{Pomeau}},
  \bibinfo{journal}{Phys. Rev. Lett.} \textbf{\bibinfo{volume}{56}},
  \bibinfo{pages}{1505} (\bibinfo{year}{1986}).

\bibitem[{\citenamefont{Brito et~al.}(1991)\citenamefont{Brito, Ernst, and
  Kirkpatrick}}]{Brito:1991JSP}
\bibinfo{author}{\bibfnamefont{R.}~\bibnamefont{Brito}},
  \bibinfo{author}{\bibfnamefont{M.~H.} \bibnamefont{Ernst}}, \bibnamefont{and}
  \bibinfo{author}{\bibfnamefont{T.~R.} \bibnamefont{Kirkpatrick}},
  \bibinfo{journal}{J. Stat. Phys.} \textbf{\bibinfo{volume}{62}},
  \bibinfo{pages}{283} (\bibinfo{year}{1991}).

\bibitem[{\citenamefont{Das and Ernst}(1992)}]{Das:1992PA}
\bibinfo{author}{\bibfnamefont{S.~P.} \bibnamefont{Das}} \bibnamefont{and}
  \bibinfo{author}{\bibfnamefont{M.~H.} \bibnamefont{Ernst}},
  \bibinfo{journal}{Physica A} \textbf{\bibinfo{volume}{187}},
  \bibinfo{pages}{191} (\bibinfo{year}{1992}).

\bibitem[{\citenamefont{Ernst and Das}(1992)}]{Ernst:1992JSP}
\bibinfo{author}{\bibfnamefont{M.~H.} \bibnamefont{Ernst}} \bibnamefont{and}
  \bibinfo{author}{\bibfnamefont{S.~P.} \bibnamefont{Das}},
  \bibinfo{journal}{J. Stat. Phys.} \textbf{\bibinfo{volume}{66}},
  \bibinfo{pages}{465} (\bibinfo{year}{1992}).

\bibitem[{\citenamefont{Grosfils et~al.}(1992)\citenamefont{Grosfils, Boon, and
  Lallemand}}]{GBL92}
\bibinfo{author}{\bibfnamefont{P.}~\bibnamefont{Grosfils}},
  \bibinfo{author}{\bibfnamefont{J.-P.} \bibnamefont{Boon}}, \bibnamefont{and}
  \bibinfo{author}{\bibfnamefont{P.}~\bibnamefont{Lallemand}},
  \bibinfo{journal}{Phys. Rev. Lett.} \textbf{\bibinfo{volume}{68}},
  \bibinfo{pages}{1077} (\bibinfo{year}{1992}).

\bibitem[{\citenamefont{Rothman}(1994)}]{Rothman:1994RMP}
\bibinfo{author}{\bibfnamefont{D.~H.} \bibnamefont{Rothman}},
  \bibinfo{journal}{Rev. Mod. Phys.} \textbf{\bibinfo{volume}{66}},
  \bibinfo{pages}{1417} (\bibinfo{year}{1994}).

\bibitem[{\citenamefont{Chen and Matthaeus}(1987)}]{Chen:1987PF}
\bibinfo{author}{\bibfnamefont{H.}~\bibnamefont{Chen}} \bibnamefont{and}
  \bibinfo{author}{\bibfnamefont{W.~H.} \bibnamefont{Matthaeus}},
  \bibinfo{journal}{Phys. Fluids} \textbf{\bibinfo{volume}{30}},
  \bibinfo{pages}{1235} (\bibinfo{year}{1987}).

\bibitem[{\citenamefont{Dab et~al.}(1991)\citenamefont{Dab, Boon, and
  Li}}]{Dab:1991PRL}
\bibinfo{author}{\bibfnamefont{D.}~\bibnamefont{Dab}},
  \bibinfo{author}{\bibfnamefont{J.-P.} \bibnamefont{Boon}}, \bibnamefont{and}
  \bibinfo{author}{\bibfnamefont{Y.-X.} \bibnamefont{Li}},
  \bibinfo{journal}{Phys. Rev. Lett.} \textbf{\bibinfo{volume}{66}},
  \bibinfo{pages}{2535} (\bibinfo{year}{1991}).

\bibitem[{\citenamefont{Kapral et~al.}(1991)\citenamefont{Kapral, Lawniczak,
  and Masiar}}]{Kapral:1991PRL}
\bibinfo{author}{\bibfnamefont{R.}~\bibnamefont{Kapral}},
  \bibinfo{author}{\bibfnamefont{A.}~\bibnamefont{Lawniczak}},
  \bibnamefont{and} \bibinfo{author}{\bibfnamefont{P.}~\bibnamefont{Masiar}},
  \bibinfo{journal}{Phys. Rev. Lett.} \textbf{\bibinfo{volume}{66}},
  \bibinfo{pages}{2539} (\bibinfo{year}{1991}).

\bibitem[{\citenamefont{Blaak and Dubbeldam}(2001)}]{Blaak:2001PRE}
\bibinfo{author}{\bibfnamefont{R.}~\bibnamefont{Blaak}} \bibnamefont{and}
  \bibinfo{author}{\bibfnamefont{D.}~\bibnamefont{Dubbeldam}},
  \bibinfo{journal}{Phys. Rev. E} \textbf{\bibinfo{volume}{63}},
  \bibinfo{pages}{021109} (\bibinfo{year}{2001}).

\bibitem[{\citenamefont{Ernst and Dufty}(1990)}]{Ernst:1990JSP}
\bibinfo{author}{\bibfnamefont{M.~H.} \bibnamefont{Ernst}} \bibnamefont{and}
  \bibinfo{author}{\bibfnamefont{J.~W.} \bibnamefont{Dufty}},
  \bibinfo{journal}{J. Stat. Phys.} \textbf{\bibinfo{volume}{58}},
  \bibinfo{pages}{57} (\bibinfo{year}{1990}).

\bibitem[{\citenamefont{Blaak and Dubbeldam}()}]{Blaak:2001JSP}
\bibinfo{author}{\bibfnamefont{R.}~\bibnamefont{Blaak}} \bibnamefont{and}
  \bibinfo{author}{\bibfnamefont{D.}~\bibnamefont{Dubbeldam}},
  \bibinfo{journal}{to be published}  (????).

\bibitem[{\citenamefont{Grosfils et~al.}(1993)\citenamefont{Grosfils, Boon,
  Brito, and Ernst}}]{GBL93}
\bibinfo{author}{\bibfnamefont{P.}~\bibnamefont{Grosfils}},
  \bibinfo{author}{\bibfnamefont{J.-P.} \bibnamefont{Boon}},
  \bibinfo{author}{\bibfnamefont{R.}~\bibnamefont{Brito}}, \bibnamefont{and}
  \bibinfo{author}{\bibfnamefont{M.~H.} \bibnamefont{Ernst}},
  \bibinfo{journal}{Phys. Rev. E} \textbf{\bibinfo{volume}{48}},
  \bibinfo{pages}{2655} (\bibinfo{year}{1993}).

\bibitem[{\citenamefont{Hanon and Boon}(1997)}]{Hanon}
\bibinfo{author}{\bibfnamefont{D.}~\bibnamefont{Hanon}} \bibnamefont{and}
  \bibinfo{author}{\bibfnamefont{J.~P.} \bibnamefont{Boon}},
  \bibinfo{journal}{Phys. Rev. E} \textbf{\bibinfo{volume}{56}},
  \bibinfo{pages}{6331} (\bibinfo{year}{1997}).

\bibitem[{\citenamefont{R\'esibois and {de Leener}}(1977)}]{Resibois}
\bibinfo{author}{\bibfnamefont{P.}~\bibnamefont{R\'esibois}} \bibnamefont{and}
  \bibinfo{author}{\bibfnamefont{M.}~\bibnamefont{{de Leener}}},
  \emph{\bibinfo{title}{Classical Kinetic Theory of Fluids}}
  (\bibinfo{publisher}{Wiley}, \bibinfo{address}{New-York},
  \bibinfo{year}{1977}).

\end{thebibliography}
\end{document}